\documentstyle[preprint,aps]{revtex}
\begin{document}
\title{Variational approach to anharmonic collective motion}
\author{G.F. Bertsch$^{(a)}$, and H. Feldmeier$^{(b)}$\\}
\address{$^{(a)}$Dept. of Physics and Inst. Nuclear Theory\\
University of Washington, Seattle\\
http://www.phys.washington.edu/$\sim$bertsch\\
$^{(b)}$Gesellschaft f. Schwerionenforschung,Darmstadt\\
http://www.gsi.de/$\sim$feldm}
\maketitle
\def\bQb{\langle\Psi_\beta|Q|\Psi_\beta\rangle}
\def\bHb{\langle\Psi_\beta|H|\Psi_\beta\rangle}
\def\bra#1{\langle #1 |}
\def\ket#1{| #1 \rangle}
\def\p{\Psi}
\def\a{\alpha}
\def\b{\beta}
\def\rv{\!\vec{\,r}}
\def\ef{e_{\!f}}
\begin{abstract}
We derive large-amplitude collective equations of motion from the
variational principle for the time-dependent Schr\"odinger equation.  
These
equations reduce to the well-known diabatic formulas for 
vibrational frequencies in the small amplitude limit.
The finite amplitude expression allows departures from
harmonic behavior of giant resonances to be simply estimated.  
The relative shift of the second phonon falls with nuclear mass $A$ as
$A^{-4/3}$ in the three modes we consider: monopole,
dipole, and quadrupole. Numerically the effect is very small
in heavy nuclei, as was found with other approaches.
\end{abstract}

\def\be{\begin{equation}}
\def\ee{\end{equation}}
\def\a{\alpha}
\def\b{\beta}
\def\eq#1{eq. (\ref{#1})}
\section{Introduction}

There has been recent interest in the harmonicity of collective motion,
with new experimental data on double-phonon excitations of the
giant dipole resonance \cite{sc93,ri93,bo96}. 
Time-dependent mean field theory provides a useful tool to study this
topic, and a number of calculations have been reported
\cite{la90,be92,ab92,po94,po96,ri96,ku96,vo95,ba92,sc90,ca89,ni95}.  Since
the calculations in the full mean-field theory are rather opaque, it 
may be of some interest to find simple approximations that 
contain the same physics.  The resulting equations of motion are quite 
intuitive
when expressed in Hamiltonian form.
With the nonlinear equation of motion we can calculate the anharmonicity
in the giant dipole resonance, relevant to the measurements of 
\cite{sc93,ri93}.

A very successful general procedure to find approximations is to employ
to the variational principle for
the Schr\"odinger equation  \cite{kr81},
\be
\label{vp}
\delta  \int dt \langle \Psi| i\partial_t-H|\Psi\rangle = 0.
\ee
This is of the form of the Lagrangian variational principle, with
the integrand playing the role of the
Lagrangian. 
The variational principle has been  attributed to Dirac; it gives a natural way to 
derive the
time-dependent Hartree-Fock approximation\cite{ke76}, and has been widely 
employed in that connection.  It also produces useful equations
of motion with more restricted assumptions about the wave function.
Morgenstern and N\"orenberg \cite{mo89} put collective motion into
the dynamics by adding variational fields corresponding to 
possible velocity potentials and the corresponding displacement fields.
Our treatment is a special case of theirs in which only two
coordinates are kept, the minimum number that
can give Hamiltonian dynamics.  

It is well known that collective motion can be generated by 
a local velocity field $Q(\rv)$ acting on a ground-state wave
function\cite{be75,ba78,br76}.
To get any kind of Hamiltonian dynamics, an additional 
degree of freedom corresponding to displacements is required.
There is a natural choice for the displacement field which
has been extensively investigated by Bohigas, et al., \cite{bo79}.
These authors studied small amplitude motion, using sum rules
to simplify the discussion.  We use the same fields and shall follow
their notation, but we are not restricted to small amplitudes.
The result will be formulas for anharmonic
effects that involve only integrals over ground state densities.
The energy shift of the second phonon excitation, $\Delta^{(2)}E$,
is calculated from the semiclassical requantization of the
Hamiltonian equations of motion.  The semiclassical requantization
has recently been favorably compared with the boson expansion approach in the
context of a solvable model\cite{kl95}. We find that this quantity is of
order $\omega_0/A^{4/3}$, where $\omega_0$ is the harmonic frequency
of the phonon and $A$ is the atomic mass number.  This strong $A$-dependence
makes the shift very small in practice.  

\section{Anharmonic collective dynamics}
  As mentioned above, the starting point is the variational principle,
eq. (\ref{vp}) above. If $\ket{\p}$ is varied in the space of Slater
determinants, the result
is time-dependent Hartree-Fock theory.    
Interesting simplified 
models are constructed by restricting $\ket{\p}$ further. 
In ref. \cite{fe90,fe95} $\ket{\p}$ is taken to be an antisymmetrized 
product of Gaussian single-particle wave functions.
In this paper we shall
assume that the motion is completely collective in the sense that
it can be generated by  a single-particle velocity
field $Q(\rv)$,
\be
\label{psi0}
\ket{\p(t=0_+)} = e^{iQ}\ket{\p_0}
\ee
Here $\ket{\p_0}$ is the ground-state wave function.  The
operator $Q$ of course acts on all the single-particle coordinates;
we shall write operators of the form  $\sum_i M(\rv_i)$ as $M(\rv)$. 
Integrating the time-dependent Schr\"odinger equation 
starting from the initial condition of \eq{psi0} gives the series
\be
\label{series}
\ket{\p(t)} = ( 1+ iQ + t [H,Q] +\cdots) \ket{\p_0}.
\ee
Our ansatz will be implemented by taking the functional form for
$\ket{\p}$ from a unitary generalization of \eq{series}.  We multiply the
fields $Q$ and $[H,Q]$ by time-dependent which
define our dynamic variables.  The operators are exponentiated to
make the transformation of the wave function unitary.  Thus
we consider a trial wave function of the form
\be
\label{ansatz}
\ket{\p_{\a\b}} = e^{i\alpha(t) Q}\ket{\p_\b} = e^{i\alpha(t) Q} e^{\beta(t) m_N[H,Q]} \ket{\p_0}
\ee
We have put in a factor of the nucleon mass $m_N$ for later convenience.
This trial wave function is a special case of eq. (2.7) of ref. \cite{mo89}
with a single field $Q$.  The two unitary transformations defined
here were first employed in ref. \cite{bo79} in treating small
amplitude collective motion.  The commutator $[H,Q]$ occurs very
frequently and we shall loosely follow the notation of ref. \cite{bo79}
with the abbreviation,
\be
\label{q1}
Q_1 \equiv m_N[H,Q].
\ee
When eq. (\ref{ansatz}) is inserted into \eq{vp}, the following
Lagrangian is obtained for the coordinates $\a$ and $\b$, 
\be 
\label{l3}
\bra{\p_{\a\b}}i\partial_t - H\ket{\p{\a\b}} =
-{\dot \alpha}\langle \Psi_{\beta}|Q|\Psi_{\beta}\rangle
-\langle \Psi_{\beta}|H|\Psi_{\beta}\rangle
-{\alpha^2\over 2 m_N}\langle \Psi_{\beta}|[Q,Q_1]|\Psi_{\beta}\rangle.
\ee
The derivation is in App. A, along with the equations of motion that
follow from the
Lagrange's equations.   We will find the phonon frequencies by 
requantizing the equations of motion, but this requires a
Hamiltonian formulation.  Thus we seek a transformation of variables 
$ (\a,\b) \rightarrow (x,p)$
together with a Hamiltonian ${\cal H}(x,p)$ such that the equations of
motion can be expressed in the form
\be
\label{heom}
\dot p = -\partial_x {\cal H}
\ee
$$
\dot x = \partial_p {\cal H}
$$
Ref. \cite{ke76} describes a procedure for obtaining
a canonical pair of variables when there are two degrees of freedom.
However, the choice of variables in  \cite{ke76} is inconvenient in that
it does not produce a quadratic Hamiltonian in the momentum.  From \eq{l3}
it is easy to see that the choice
$ x = 
 \langle\Psi_\beta|Q|\Psi_\beta\rangle,\,\, p = \a$ gives a canonical pair 
$(x,p)$ with a simple kinetic energy term in the Hamiltonian.  However, the
Hamiltonian is difficult to express explicitly in terms of this $x$.
Therefore we use instead the canonical pair $(\b,p_\b)$ with
$$
p_\b = \a \bra{\p_\b}[Q,Q_1]\ket{\p_\b}
$$
which allows us to keep $\b$ as the coordinate variable\footnote{Since
both are canonical pairs it can be shown that $pdx = p_\b d\b$ so that the
phase integral (Appendix B) is invariant.}.
The Hamiltonian in this representation may be written as
\be
\label{ham}
{\cal H}(\b,p_\b) = {p_\b^2\over 2 m(\b)} + U(\b)
\ee
with 
\be
\label{mass}
 m(\b) = m_N \langle\Psi_\beta|[Q,Q_1]|\Psi_\beta\rangle
\ee
$$
U(\b) = \langle\Psi_\beta|H|\Psi_\beta\rangle
$$
The reader may verify that the Hamiltonian equations (\ref{heom})
defined this way are equivalent to \eq{eom1a} and (\ref{eom2a}).
 
     To make use of use of the equations of motion, we need evaluate 
$\b$-dependent matrix elements appearing in \eq{mass}. 
This can be done in two different
ways, depending on whether we apply the operator $\exp(\b Q_1)$ to the
left or to the right. The first way is to explicitly apply the
operator on the wave function.
We shall only consider nonrelativistic Hamiltonians with local potentials, 
so the operator
$Q_1=m_N[H,Q]$ is a first order derivative:
\be
Q_1 = -{\nabla^2 Q\over 2 } - {\nabla Q } \cdot \nabla
\ee
The exponentiated operator $\exp(\beta Q_1)$ acting on a function
of a single coordinate variable, such as $x$, may be expressed in closed form as follows
\be
e^{\beta Q_1} f(x,y,z) = \sqrt{ \partial_x Q(x')\over
\partial_x Q(x)} \: f(x',y,z).
\ee
The displaced coordinate $x'$ is 
obtained by integrating from $x'$ a distance $x-x'$ that 
satisfies
\be
\beta = \int_x^{x'} {ds \over \partial_x Q(s) }.
\ee 
The derivation of this formula may be found in ref. \cite{sc96}.

The other way to calculate matrix elements is to apply  the
unitary transformation to the operator being evaluated.  
Thus we use the identity
\be \label{optrans}
\langle \Psi_\beta | {\cal M} | \Psi_\beta\rangle
= \langle \Psi_0 | \Big( e^{-\beta Q_1} {\cal M} e^{ \beta Q_1} \Big)
| \Psi_0\rangle.
\ee
This will turn out to be very convenient for matrix elements of
scaling displacements.

The special case where
Q is a quadratic function of the coordinates gives a particularly
simple form for the transformed coordinates \cite{bo79}, namely
they are scaled by a factor.  For example, for the field $Q=x^2/2$,
the transformation is 
\be
e^{\beta Q_1} (x,y,z) e^{-\beta Q_1} = (x',y',z')=(e^{-\beta} x,y,z)
\ee
and the wave function $\Psi_\beta$ is given by
\be \label{wavetrans}
\Psi_\beta(x,y,z) =e^{-\b/2}\Psi_0(e^{-\beta} x, y,z)
\ee

\section{A Model Hamiltonian}
We wish to apply the equations of motion to a variety of giant
resonances, and will need a detailed model for $H$ in order to
construct $U(\b)= \bHb $.  A good balance between simplicity and
realism is provided by the Skyrme-like form for the Hamiltonian
density,
\be
\label{skyrme}
h = \rho_0 \left[\,\tau +  v_a n^2 +  v_b  n^{7/3}\,\right]
\ee
where $\tau$ is the kinetic energy density and
$n$ the density, both in units of nuclear matter density $\rho_0$.
The coefficients $v_a$ and $v_b$ are 
determined to reproduce nuclear saturation density $\rho_0 \approx 0.16 $ 
fm$^{-3}$ (with 
corresponding Fermi energy $\ef\approx 36$ MeV) and the
nuclear matter binding energy  $B\approx 16$ MeV per nucleon.
If we express $n$ in units of the saturation density the 
parameters may be expressed
\be
\label{vavb}
v_a = -4 B - {6\over 5} \ef \approx -107 {\rm \,\,MeV}
\ee
$$
v_b = 3 B + {3\over 5} \ef \approx 70 {\rm \,\,MeV}.
$$
The power dependence of the third term in eq. (\ref{skyrme}) is 
not obvious.
{From} many-body theory, one expects the
energy of a dilute Fermi gas to be a series in 
powers of $k_f$ or $n^{1/3}$.  Eq.
(\ref{skyrme}) thus represents the first three terms of that series.
The parameterization $n^{7/3}$ for the
third term also predicts a compressibility not far from that
required by the empirical monopole systematics.  It should be mentioned
that \eq{skyrme}
lacks momentum-dependent interactions, which
are certainly present in the empirical single-particle Hamiltonian.

For treating the giant dipole resonance, we also need to know the isospin
dependence of $H$.  The kinetic energy has an obvious isospin
dependence arising from the separate Fermi energies of neutrons and
protons.  We shall add isospin-dependent potential energy terms with
the same density-dependence as in eq. (\ref{skyrme}), and 
require that the 
isospin dependence of the semiempirical mass formula to be reproduced.
The binding energy per particle $\epsilon(n,n_\tau) = h/(n\rho_0)$
in the Fermi gas approximation
is expressed as follows, with proton and neutron densities
written $n_p = n/2 + n_\tau$ and $ n_n = n/2 - n_\tau$.\\
\be
\label{extended}
\epsilon(n,n_\tau) = {3\over 5} \ef {n^{2/3}\over 2} \left[
         \left(1 + {2 n_\tau\over n}\right)^{5/3}
       + \left(1 - {2 n_\tau\over n}\right)^{5/3} \,  \right] 
       + v_a n + v_b n^{4/3} +v_{\tau} n_\tau^2/n 
\ee
Expanding this in powers of $n_\tau$, we have
\be
\epsilon(n,n_\tau) \approx \epsilon(n,0) + 
          \left( {4 \over 3} e_{\!f}  n^{2/3}+v_\tau n \right)
          \left({n_\tau\over n}\right)^2.
\ee 
The semiempirical mass formula has isospin dependent terms,
\be
B(A,Z) =B(A,A/2)+ b_{sym} {(N-Z)^2\over A^2} + b_c {Z^2\over A^{4/3}}+\cdots
\ee
with $b_{sym}\approx 25$ MeV.  Assuming neutrons and protons
occupy the same volume, $n_\tau/n = (Z-N)/2A$ and we may
relate the coefficient in \eq{extended} to $b_{sym}$ as
\be
{4 \ef \over 3} + v_\tau  = 4b_{sym}
\ee
Putting in the Fermi energy $\ef \approx 36$ 
MeV, we find numerically
\be
\label{vtau}
v_\tau  \approx 50 {\rm\,MeV}.
\ee

\section{Finite amplitude excitations}

We now treat the excitation of the giant monopole, dipole and quadrupole
modes of vibration.
For each multipole, we will define a 
collective field $Q$, and then evaluate the harmonic frequency and
the nonlinear corrections.  The functions that are required for
this are the expectation of the Hamiltonian in the $\b$-deformed
state, which we expand as
\be
\label{useries}
U(\b) = {k\over2}\b^2+{k_3\over3}\b^3+{k_4\over4}\b^4+\cdots
\ee
We have defined the energy scale so that $U(0)=0$.  The 
linear term in the expansion vanishes because of the stability
of the ground state.  We also need to expand the inertia to
second order in $\b$.  We write this as
\be
\label{mseries}
 m(\b) = m_N \bra{\p_\beta} [Q,Q_1]\ket{\p_\beta} =
m(1+m_1\b +m_2 \b^2+\cdots).
\ee
The key formulas are the equation for the frequency in the
harmonic limit,
$$
\omega_0 = \sqrt{k\over m}
$$
and the formula for the energy shift of the second phonon.  It
is convenient to express this in terms of an energy parameter
$E_{anh}$ as 
\be 
\label{diff2}
\Delta^{(2)}E= E_2 -2 E_1+E_0 =
2 {\omega_0^2\over E_{anh}}.
\ee
The derivation of the expression for $E_{anh}$ in terms of the
nonlinear coefficients $k_3,k_4,m_1$ and $m_2$ is given in
App. B.  The result is
\be
\label{eah1}
E_{anh}^{-1} = {5\over12}{k_3^2\over k^3} -{3 \over 8}{k_4\over  k^2}
-{k_3 m_1\over 4 k^2 } -{m_1^2\over 16k}+ {m_2\over 4 k  }.
\ee
All the $k$'s in this equation scale with mass number as $k_i\sim A$,
while the $m_i$ are independent of $A$ in the droplet limit.  We thus
see that the anharmonicity 
energy scale $E_{anh}$ varies with mass number as
$$
E_{anh} \sim A.
$$
We now consider the various multipoles in turn, starting with 
the isoscalar monopole and
quadrupole modes.

\subsection{Monopole}
The monopole field $Q$ for a uniform sphere with a sharp edge
would be
proportional to the $j_0$ spherical Bessel function, but
in practice the nuclear surface cannot be ignored even for
large nuclei.  The compressibility is less in the surface,
and this has the consequence that the velocity potential
is more like the simple scaling form \cite{bl95},
\be
Q = r^2/2
\ee
We shall construct the nonlinear dynamics with this field.
It is most convenient to apply the transformation to
the Hamiltonian in this case.  The inertia is
\be 
m(\b)= m_N\bra{\p_\beta} [Q,Q_1]\ket{\p_\beta} 
= e^{2\b} m_N \bra{\p_0}r^2\ket{\p_0}= e^{2\b}m_N A \langle r^2\rangle, 
\ee
where we $\langle r^2\rangle$ denotes the mean square radius of the
ground state.
The expectation value of the Hamiltonian is
\be
\label{U-m}
\bra{\p_\beta}H\ket{\p_\beta} = \rho_0\left[
  e^{-2\b}     \int d^3r \, \tau_0(\rv) 
+ e^{-3\b} v_a \int d^3r \, n_0^2(\rv)
+ e^{-4\b} v_b \int d^3r \, n_0^{7/3}(\rv) \right]
\ee
where $\tau_0(\rv)$ denotes the kinetic energy density of the 
ground state $\ket{\p_0}$ and $n_0(\rv)$ the particle density.  This
formula is derived using the relations \eq{optrans} and \eq{wavetrans},
which yield for the monopole field
$$
e^{-\b Q_1}\tau_0(\rv)e^{\b Q_1} = e^{-5\b}\tau_0(e^{-\b}\rv)
$$
and 
$$
e^{-\b Q_1} n_0(\rv) e^{\b Q_1}  = e^{-3\b} n_0(e^{-\b}\rv).
$$
We next expand \eq{U-m} in the power series in $\b$.  The linear
term vanishes because of the saturation condition, \eq{vavb}.  The
quadratic term, giving the effective restoring force, 
is 
\be
k = \partial^2_\b\bra{\p_\b}H\ket{\p_\b}|_{\b=0} 
=\rho_0
\left[\, 4\int d^3 r\,\tau_0(\rv)
+9 v_a \int d^3 r\, n_0^2(\rv)
+16 v_b\int d^3 r\, n_0^{7/3}(\rv) \, \right]\ . 
\ee
For a spherical drop with radius $R$, this is equal to $A$ time the 
nuclear matter compressibility $K$ 
if one can make the large-$A$ approximation $ n_0(\rv) = \theta(R-r)$
which yields 
$\rho_0\int\!d^3r \, n_0^2(\rv) = \rho_0\int\! d^3 r\,n_0^{7/3}(\rv)= A$ 
and 
$\bra{\p_0}-\nabla^2/2 m_N\ket{\p_0} = \rho_0\int\! d^3r \,
\tau_0(\rv) = 3 \ef A/5$. 
Surface
effects of course spoil this approximation, and the fact of the
matter is that they have an exaggerated importance because 
the $v_a$ and $v_b$ are separately large with opposite sign.  However,
for our purposes it is an unnecessary refinement to improve on the
nuclear matter approximation.  
The harmonic
approximation for the frequency is then the well-known collective
formula,
\be
\omega_0^2 = {K \over m_N \langle r^2\rangle}
\ee 

The nonlinear coefficients in the expansion of $U$ and $m(\b)$ 
are given numerically
in Table~I.  Combining these according to \eq{eah1} we obtain for the
anharmonicity parameter
$$
E_{anh} = 40  A MeV
$$
This is more than a factor of $A$ larger than the vibrational frequency,
implying that the shift will be very small.  
For example, for $^{208}$Pb the shift from \eq{diff} is
$\Delta^{(2)} E = 0.05$ MeV.  This of course is completely
insignificant as a measurable effect.

\subsection{Giant Quadrupole}
The theory of the giant quadrupole anharmonicity is very 
similar.
We define the  isoscalar quadrupole field as
\be 
Q = z^2-{1\over2} (x^2+y^2)
\ee
and hence the wave function transforms as (see \eq{wavetrans})
$$
\p_\b(x,y,z) = e^{\b Q_1} \p_0(x,y,z) = \p_0(e^\b x,e^\b y, e^{-2\b} z).
$$
To evaluate matrix elements of various fields, we shall
assume that the ground state of the nucleus is spherical.  Then
the inertia is given by 
\be
\bra{\p_\b}[Q,Q_1]\ket{\p_\b} = 2e^{2\b}m_N\bra{\p_0} r^2\ket{\p_0}
\approx 2Am_N e^{2\b}\langle r^2\rangle 
\ee
and the collective potential energy in the Hamiltonian is
\be
\bra{\p_\b} H \ket{\p_\b} = \Big({1\over 3}e^{-4\b}+{2\over 3}
e^{2\b}\Big)\bra{\p_0}{-\nabla^2\over 2
m_N }\ket{\p_0}
\ee 
$$
\approx \Big(e^{-4\b}+2e^{2\b}\Big) {\ef \over 5} A.
$$
In the last step we have used the Fermi gas estimate for the kinetic
energy.  Expanding these as power series in $\b$, we obtain the 
coefficients in Tables I and  II.
The harmonic limit is given by 
the simple formula 
\be 
\omega_0^2 = {k\over m} = { 12  \ef \over 5 m_N \langle r^2\rangle }.
\ee 

The power series expansion to higher order is also rather simple for
the quadrupole, since the only nuclear parameters that enter are
$A,\ef,$ and $\langle r^2\rangle$. 
It turns out
that there are strong cancellations among the different terms in 
\eq{eah1}, giving for the anharmonicity parameter
\be
E_{anh} = {288\over 25} A \ef.
\ee
This is larger than the parameter for the monopole, implying that
the shift would be even smaller.

\subsection{Giant dipole resonance}
The field $Q$ for the giant dipole resonance is not 
as simple as the other cases. In light nuclei, the energetics of the
giant dipole state suggests that the $Q$ is close to being
the simple operator  $\tau_z z $ but this form
gives the wrong $A$-dependence to describe the dipole energies
in heavy nuclei.  The Steinwedel-Jensen model takes an
opposite extreme, positing that the displacement field
vanishes at the nuclear surface.  This can be generated by
a  velocity field such as 
$$
Q= \tau_z z (1-r^2/3R^2).
$$
where $R$ is the nuclear radius.
We will adopt this form to investigate the nonlinearity, although
the model predicts too high a frequency for the dipole.
The displacement field then has the form
$$
Q_1 = {x^2+y^2+3z^2-3R^2\over 3R^2}\partial_z
+{5 z \over 3 R^2} +{2 z\over 3 R^2}\Big( x\partial_x+y\partial_y\Big).
$$
The coordinate $\b$ associated with this $Q_1$ has the dimensions
of length.  Unlike in the monopole and quadrupole cases, we were unable
to find an analytic form for the needed expectation values.  This is
due to the mixing of the Cartesian coordinates in
$Q_1$.  Instead, we use the expansion \eq{expand}
explicitly to evaluate the inertia and the kinetic energy term
of the Hamiltonian.  For the inertia, we must expand to second
order as
$$
\bra{\p_\b}[Q,Q_1]\ket{\p_\b} = 
\bra{\p_\b}[Q,Q_1]\ket{\p_\b} -\b\bra{\p_\b}[Q_1,[Q,Q_1]]\ket{\p_\b}
+{\b^2\over 2}\bra{\p_\b}[Q_1,[Q_1,[Q,Q_1]]]\ket{\p_\b} 
$$
We shall evaluate this approximating the density as that of a 
uniform drop.  The result is
$$
m = A m_N \Big( {32\over 63} - {2944\over 5103}  {\b^2\over R^2}+\cdots\Big)
$$
Notice that $\b$ appears in the nonlinear terms in the dimensionless
ratio $\b/R$.   

  The kinetic energy operator is
treated by applying \eq{expand} to each of the two gradients that
it contains.  The algebra is very tedious, and we have used
a computer program for the manipulations.  
We quote here as an example, the 
$x$ gradients expanded to second order in $\b$,
$$
\partial_x' = \partial_x + {2\b\over 3 R^2} \Big(z \partial_x+
x\partial_z\Big) + {\b^2\over 9 R^2}\Big(5x+(3 R^2+x^2-y^2-z^2)
\partial_x + 2xy\partial_y +6 xz\partial_z\Big).
$$
We actually
need the gradient evaluated to fourth order, but it does not seem
useful to display the full expression.  We evaluate
the kinetic energy in the Fermi gas approximation by applying the
transformed gradient to a plane wave state, taking the modulus
squared, and integrating over a spherical Fermi surface.  For
each term in $\b$, we keep only the highest power of $R$.  This
is 
$$
\bra{\p_\b}{-\nabla^2\over 2 m}\ket{\p_\b}=
           {3\over 5} \ef A + {44\over 45} \ef A {\b^2\over R^2}+
           {168853\over 212625} \ef A {\b^4\over R^4}+\cdots
$$

It still remains to evaluate the potential energy.  We found this
easier to do by applying the transformation to the
wave function, and explicitly constructing the transformed
single-particle density to the needed order in $\b$.  We will
again approximate the ground state as a uniform sphere, which
means that gradients of the wave function are ignored in
evaluating the effect of the operator.  Of course, the 
wave function gradients cannot be ignored in the surface region,
but we have constructed the operator $Q$ to have no surface
contributions.  We only need the wave function to 
third order in $\b$ to  evaluate the potential energy to fourth order.
The wave function is
$$
\ket{\p_\b} = \Big( 1 + \b Q_1 + {\b^2\over 2} Q_1^2 + {\b^3\over 6}
Q_1^3 + \cdots\Big) \ket{\p_0}
$$
Calculating the density from this, we see that both the isovector
and isoscalar densities are affected by the transformation.  The
densities are given to order $\b^2$ as
$$
n_\b(\rv) =1 + x{\b^2\over R^2} + \cdots
$$
$$
n_{\tau\b}(\rv) = {100 z^2 \b^2\over 9 R^4}+\cdots
$$
Inserting the densities to order $\b^4$ in the potential energy 
function, we find the following expression for the potential energy
$$
\bra{\p_\b}v\ket{\p_\b} = \bra{\p_0}v\ket{\p_0}+ {20 \over 9}v_\tau A 
{\b^2\over R^2}  - {1040\over 1701} v_\tau A {\b^4\over R^4} + \cdots
$$

Notice that the coefficients of $\b^n$ are all of the form
of a constant times $A/R^n$.  This implies that $E_{anh}$ will
depend on nuclear size as $A$.  There is a cancelation between
the kinetic and potential contributions to
$k_4$, so the resulting anharmonicity is very small,
$$
|E_{anh}|   > 100     A {\rm\,\,MeV}
$$
giving once more a negligible energy shift for the double excitation.

\section{Conclusion}

Our conclusion, that anharmonic effects are extremely small in giant
vibrations, is in agreement with earlier studies. However, the precise
$A$-dependence had not been made clear.    For example, in ref. \cite{po96},
the authors expected $\Delta^{(2)} E$ to scale with $A$ as $\Delta^{(2)} 
E\sim A^{-2/3}$.  See also \cite{bm75}.  This implies that 
$E_{anh}$ would be independent of $A$, disagreeing with our
linear $A$ dependence.

Our model could be improved in a number of ways.  The actual field
for the dipole has a considerable amount of surface displacement, even
for heavy nuclei, and a more realistic field could be employed.
The Hamiltonian should include momentum-dependent interactions to
be more realistic.  However, there is no reason to think these
improvements would change the picture in a qualitative way, and
it hardly seems worthwhile to calculate the very small effect 
more accurately.

More interesting and challenging is to develop a nonlinear collective
description of the low collective modes, in particular the
octupole vibration. Part of the doubly excited octupole has 
recently been identified in $^{208}$Pb (ref. \cite{ye96}), 
but the only theory up to now is
the rather opaque second RPA approximation.  To use our treatment,
one would have to find a nearly local operator which would generate
the low octupole, i.e. a field for which the associated sum rule
would be nearly exhausted by the low mode.
\section{Acknowledgment}
We would like to thank R. Broglia, W. N\"orenberg and J. Schnack 
for discussions.
G.F. Bertsch acknowledges support as a Humboldt Foundation
Senior Fellow and also support by the Department of Energy
under Grant DE-FG06-90ER40561.  H. Feldmeier acknowledges
the hospitality and financial support of the Institute for Nuclear Theory 
during his stay there.
\appendix

\section{Derivation of equation of motion}
   We shall evaluate the variational principle, eq. (1), using as
a trial wave function \eq{ansatz}, $\ket{\p_{\a\b}} = \exp(i\alpha Q)
\exp(\beta m_N[H,Q]) \ket{\p_0} = \exp(i\a Q) \ket{\p_\b}$.  The variational 
principle will be used
to determine the coefficients $\alpha$ and $\beta$; $Q$ is some local function 
of position and 
$\ket{\p_0}$ is the ground state.
The variation with respect to
$\alpha$ and $\beta$ in eq. (1) yields the usual Lagrangian equations of 
motion,
\be
\label{l1}
{d \over dt} {\partial {\cal L} \over \partial {\dot \alpha}} -
{\partial {\cal L} \over \partial \alpha} =0
\ee
\be
\label{l2}
{d \over dt} {\partial {\cal L} \over \partial {\dot \beta}} -
{\partial {\cal L} \over \partial \beta} =0
\ee
where ${\cal L}$ is the integrand in eq. (1).
\be
\label{lag}
{\cal L} = 
\langle\Psi_{\alpha\beta}|i\partial_t|\Psi_{\alpha\beta}\rangle
-\langle \Psi_{\alpha\beta}|H|\Psi_{\alpha\beta}\rangle
\ee
{From} this point it is simply algebra to reduce eq. (A1-A3) to a more
transparent form.  We shall use an identity for expanding the
operator product $e^{-A} B e^A$ in commutators,
\be
\label{expand}
e^{-A} B e^A = B + [B,A] + {1\over 2} [[B,A],A] + {1 \over 3
!}[[[B,A],A],A]+\cdots
\ee
We also need two corollary identities,  
\be
\label{c1}
\partial_\alpha (e^{-\alpha A} B e^{\alpha A}) = e^{-\alpha A} [B,A] e^{\alpha A} 
\ee
and
\be
\label{c2}
e^{-A} A e^A = A.
\ee

We begin by examining the second  term in eq. (\ref{lag}),
$\langle \Psi_{\alpha\beta}|H|\Psi_{\alpha\beta}\rangle$.  The
$\alpha$ dependence is made more explicit by expanding
$e^{-i\alpha Q} H e^{i\alpha Q}$ in commutators.  If $Q$ is
local (and $H$ is quadratic in the momenta), the third and higher-order
commutators vanish.  From eq. (\ref{expand}) and \eq{q1} the expansion is
\be
\langle \Psi_{\alpha\beta}|H|\Psi_{\alpha\beta}\rangle=
\langle \Psi_{\beta}|H|\Psi_{\beta}\rangle
+i{\alpha\over m_N} \langle \Psi_\beta|Q_1|\Psi_\beta\rangle
+{\alpha^2\over 2 m_N}\langle \Psi_{\beta}|[Q,Q_1]|\Psi_{\beta}\rangle.
\ee
Next we argue that the middle term in this sum vanishes.
First observe that by \eq{c2} $\langle \Psi_\beta|Q_1|\Psi_\beta\rangle=
\langle \Psi_0|Q_1|\Psi_0\rangle$ does not depend on $\b$. The fact that
$\ket{\p_0}$ is a 
stationary state of the Hamiltonian implies 
\be
\label{q1eq0}   
0= \partial_\a\bra{\p_{\a,\b=0}}H\ket{\p_{\a,\b=0}}\bigg|_{\a=0} = {i\over m_N}
\bra{\p_0} Q_1\ket{\p_0} = {i\over m_N} \bra{\p_\b}Q_1\ket{\p_\b}.
\ee
The time derivative in
eq. (\ref{lag}) contains two terms, 
$$
\langle\Psi_{\alpha\beta}|i\partial_t|\Psi_{\alpha\beta}\rangle
=-{\dot \alpha}\langle \Psi_{\beta}|Q|\Psi_{\beta}\rangle
+i {\dot \beta}\langle \Psi_\beta|Q_1|\Psi_\beta\rangle 
$$
The second term vanishes by \eq{q1eq0}.
The complete expression for the Lagrangian
then becomes 
\be
\label{lag2}
{\cal L} =  
-{\dot \alpha}\langle \Psi_{\beta}|Q|\Psi_{\beta}\rangle
-\langle \Psi_{\beta}|H|\Psi_{\beta}\rangle
-{\alpha^2\over 2 m_N}\langle \Psi_{\beta}|[Q,Q_1]|\Psi_{\beta}\rangle.
\ee
The dependence on $\alpha$ is now entirely explicit.  Inserting
${\cal L}$ in eq. (\ref{l1}) yields
\be
-{\partial_t} \langle \Psi_{\beta}|Q|\Psi_{\beta}\rangle
+{\alpha\over m_N} \langle \Psi_{\beta}|[Q,Q_1]|\Psi_{\beta}\rangle
=0
\ee
This is simplified with the help of eq. (\ref{c1}) to
\be
\label{eom1a}
\dot \beta ={\alpha\over m_N} 
\ee
We next carry out the derivatives in the second Lagrangian equation, 
eq. (\ref{l2}), to obtain
\be
\label{eom2a}
\dot\alpha\partial_\beta \langle \Psi_{\beta}|Q|\Psi_\b\rangle
+\partial_\beta\langle \Psi_{\beta}|H|\Psi_\b\rangle
+{\a^2\over 2 m_N}\partial_\b\langle \Psi_\b|[Q,Q_1]|\Psi_\b\rangle=0.
\ee
This is as far as we can simplify it without approximation beyond the
basic ansatz, \eq{ansatz}.  The
harmonic limit is obtained by using the expansion \eq{expand} and by keeping 
in \eq{eom2a} only linear terms in
$\a$ and $\b$.  By the stationarity of the ground state 
$\partial_\b \bra{\p_\b}H\ket{\p_\b}|_{\b=0} = 0 $ and the equation of
motion reduces to
\be 
\label{eom2b}
\dot \alpha\langle \Psi_0|[Q,Q_1]|\Psi_{0}\rangle+
\beta     \langle \Psi_0|[[H,Q_1],Q_1]|\Psi_{0}\rangle=0.
\ee
The frequency of oscillation\cite{bo79} is then given by inserting \eq{eom1a}
into \eq{eom2b} as
\be
\omega^2 = {  \langle \Psi_0|[[H,Q_1],Q_1]|\Psi_{0}\rangle
\over m_N\langle \Psi_0|[Q,Q_1]|\Psi_{0}\rangle}={M_3\over M_1},
\ee
where $M_n$ is the $n$-th energy moment of the transition strength,
$$
M_n = \sum_i \langle 0|Q|i\rangle^2 (E_i-E_0)^n.
$$

\section{Anharmonicity}
In this appendix we derive the frequency shift of multiple
phonon excitations due to the anharmonicity of the 
equation of motion (\ref{eom2a}).  Our derivation proceeds through
the Bohr-Sommerfeld quantization of the classical orbits.
This requires the phase integral
$$
\phi = \int p dx
$$
using a Hamiltonian representation of the equations of motion. 
The condition that an eigenstate be at energy $E$ is that
classical phase accumulated over the orbit be an integral multiple of
$2\pi$.
Taking the form \eq{ham} for the Hamiltonian, 
the energy $E_n$ of the $n$-th state
satisfies
\be
\label{bohr}
\phi(E_n)=
 \int_{\b_1}^{\b_2} \sqrt{2m(\b)\Big(E_n-U(\b)\Big)} d \b = n \pi  
\,\,\,\,\,n=0,1,2,\cdots
\ee
where $\b_{1,2}$ are the classical turning points given by 
$U(\b_{1,2})=E_n$
\footnote{If the additional phase of $\pi/4$ is added for each
turning point, \protect{\eq{bohr}} 
gives the exact energies in the harmonic limit.}.

We now assume that the anharmonicity is weak, so that all quantities
can be expanded in power series in $\b$, which we wrote as
\eq{useries} and (\ref{mseries}).  
To evaluate the integral eq. (\ref{bohr}) to a given order in $\beta$, we
change variable to make the energy difference under the square root
a simple quadratic function.  Defining a variable
$ z = \sqrt{U(\b)/E}$, we
write the integral as
\be
\phi(E) = \sqrt{2E}\int_{-1}^1 \sqrt{1-z^2} \sqrt{m(\b)} {d\b\over d z} d z.
\ee 
The second square root and derivative are expanded in powers of $\beta$.
The latter is obtained via the series for $\beta(z)$,
\be 
\beta = \sqrt{2E\over k} (z - {2^{1/2}k_3\over 3  k^{3/2}} E^{1/2}z^2+
({5\over9} {k_3^2\over k^3}-{k_4\over 2 k^2}\Big) E z^3 + \cdots), 
\ee
with the derivative
$$
d \beta = \sqrt{2E\over k} (1 - 
2^{3/2}{k_3\over 3  k^{3/2}} E^{1/2}z
+
({5\over3} {k_3^2\over k^3}-{3 k_4\over 2 k^2}\Big) E z^2 +\dots ) dz.
$$
The integrals are then elementary to evaluate, giving for $\phi$,
\be 
\label{phi}
\phi = \pi {E\over \omega_0}
+\pi\Big({5\over12}{k_3^2\over k^3} -{3 \over 8}{k_4\over  k^2}
-{k_3 m_1\over4 k^2 } -{m_1^2\over16 k}+ {m_2\over4 k }\Big){E^2\over\omega_0}+\cdots
\ee
where $\omega_0=\sqrt{k/m}$ is the small amplitude harmonic frequency.
The anharmonicity in this limit is controlled by the combination of 
nonlinear coefficients in parentheses. It has the
dimensions of inverse energy and we shall abbreviate it as
\be
\label{eah}
E_{anh}^{-1} = {5\over12}{k_3^2\over k^3} -{3 \over 8}{k_4\over  k^2}
-{k_3 m_1\over 4 k^2 } -{m_1^2\over 16 k}+ {m_2\over 4 k  }.
\ee
To find the energy shifts, we insert \eq{phi} in \eq{bohr}  
and invert the resulting power series 
that expresses  $n$
in terms of $E$.  The result is
\be
\label{anharm}
E_n =n \omega_0-
n^2{\omega_0^2\over E_{anh}}+\cdots
\ee
The shift of the double phonon with respect to the single
phonon excitation is of direct experimental interest.  To
leading order, this is evaluated from eq. (\ref{anharm}) as
\be 
\label{diff}
\Delta^{(2)}E= E_2 -2 E_1+E_0 =
-2 {\omega_0^2\over E_{anh}}.
\ee
Eq. (\ref{anharm}) is used in Sect. IV.  

Another way to derive the frequency shift would be to find the
classical frequencies at energies 
$\omega_0$
and $2\omega_0$.  Those energies correspond to wave packets made
of states $n=0,1$ and $n=1,2$, respectively. Thus the classical
frequencies
correspond to the energy differences $E_1-E_0$ and $E_2-E_1$,
respectively, and the shift would be calculated as 
\be
\Delta^{(2)}E= \omega(\omega1) -\omega(\omega_0)
\ee
Classical orbital perturbation theory may be used to express these
frequencies in terms of the anharmonicities $m_1,m_2,k_3$ and $k_4$.
We have verified that the two methods give the same result for the
$k_4$ dependence.

\begin{table}
\caption{Coefficients of the collective inertia 
expansion \protect{\eq{mseries}} }
\begin{tabular}{r|ccc} 
\hline
Mode  & $m/A$  & $m_1$ & $m_2$\\
\hline
Monopole  &  $m_N\langle r^2 \rangle $ &  2 & 2  \\
Quadrupole & $2m_N \langle r^2 \rangle $  & 2     &  2   \\
Dipole     &  $32 m_N/63$   &0     &     $-0.58/R^2$             \\
\end{tabular}
\end{table}
\begin{table}
\caption{Coefficients of the collective Hamiltonian
expansion \protect{\eq{useries}} }
\begin{tabular}{r|ccc} 
\hline
Mode  & $k/A$  & $k_3/A$ & $k_4/A$   \\
\hline
Monopole  &  $K =230.$ MeV   &   -867.  &   1580.     \\
Quadrupole & $24 \ef /5= 173.$ MeV      & 173.        &346.   \\
Dipole (kinetic)  & $88 \ef /45R^2$     &   0         & $3.2\ef /R^4$  \\
Dipole (potential) & $40v_\tau/9R^2$& 0  & $-2.4 v_\tau/R^4$ \\
\end{tabular}
\end{table}

\end{document}